\documentclass[twocolumn]{webofc}
\usepackage[varg]{txfonts}   % Web of Conferences font
\usepackage{hyperref}
\usepackage{url}

\newcommand{\csavg}{\langle c_s^2 \rangle}
\newcommand{\cs}{c_s^2}

\newcommand{\corr}[1]{{\color[rgb]{0,0,0}{#1}}}

\hypersetup{colorlinks=true,citecolor=blue,urlcolor=blue,linkcolor=blue}
\begin{document}
\title{Kinetic theory and the speed of sound in dense matter}
%
% subtitle is optionnal
%
%%%\subtitle{Do you have a subtitle?\\ If so, write it here}

\author{\firstname{Michał} \lastname{Marczenko}\inst{1,2}\fnsep\thanks{\email{michal.marczenko@uwr.edu.pl}}
        % etc.
}

\institute{%
Institute of Theoretical Physics,\\ University of Wroc\l{}aw, 
plac Maksa Borna 9, 50-204 Wroc\l{}aw, Poland
\and%
Incubator of Scientific Excellence - Centre for Simulations of Superdense Fluids,\\ University of Wroc\l{}aw, 
plac Maksa Borna 9, 50-204 Wroc\l{}aw, Poland
}

\abstract{We discuss a constraint on the speed of sound, $\cs$, derived from relativistic kinetic theory and show how it can be expressed in terms of the average sound speed, $\csavg$. This reformulation highlights the interplay between instantaneous and integrated stiffness of the equation of state and allows the kinetic-theory bound to be visualized as a restriction in the $\cs$-$\csavg$ plane.}
\maketitle
%

% \section{Introduction}

The speed of sound plays a central role in characterizing the stiffness of the equation of state (EOS) of dense matter~\cite{Altiparmak:2022bke}. While the local sound speed,
\begin{equation}
\cs = \frac{\partial p}{\partial \epsilon} \rm ,
\end{equation}
where $p$ is the thermodynamic pressure and $\epsilon$ is the energy density, encodes the instantaneous response of the medium, it is often useful to consider integral or averaged measures that capture the cumulative stiffness over a finite density or energy interval. The averaged quantities naturally arise in global consistency analyses and have been shown to be closely related to macroscopic observables and thermodynamic inequalities~\cite{Marczenko:2024uit, Marczenko:2025hsh}. The average sound speed
\begin{equation}\label{eq:1}
        \csavg = \frac{1}{\epsilon}\int\limits_0^\epsilon \mathrm d \epsilon' \cs(\epsilon') = \frac{p  }{\epsilon},
\end{equation}
was introduced as a way of characterizing the overall stiffness of an EOS over a finite energy-density range, rather than focusing on local extrema in the speed of sound.

% \section{Kinetic-theory bound}
A natural question is how such averaged measures are constrained by microscopic consistency conditions. Relativistic kinetic theory (KT) implies a nontrivial bound on the speed of sound~\cite{Olson:2000vx},
\begin{equation}
        \cs \le \frac{\epsilon - p/3}{\epsilon + p} = \frac{1 - p/(3\epsilon)}{1 - p/\epsilon},
\end{equation}
which follows from the requirement that the relaxation time appearing in causal transport equations remains positive and that the corresponding dispersion relations are stable. Unlike the purely causal condition $\cs \le 1$, this bound depends explicitly on the thermodynamic state of the system through the ratio of the thermodynamic pressure and energy density and therefore constrains not only signal propagation but also the admissible stiffness of the EOS.

In this contribution, we examine how the constraint motivated by relativistic kinetic theory limits the admisisble combinations of the speed of sound and its average.

% \section{Discussion}

An important observation is that this inequality can be rewritten in terms of the average speed of sound. Using Eq.~\eqref{eq:1}, the KT bound can be equivalently written as
\begin{equation}
        \cs \le \frac{1 - \langle \cs \rangle/3}{1 + \langle \cs \rangle} \rm.
\end{equation}
This representation shows that the KT constraint depends explicitly on both the speed of sound and its averaged value. As a result, even moderate values of the average sound speed can place nontrivial restrictions on the maximum locally admissible $\cs$. The KT constraint has previously been implemented in model-independent constructions of the EOS~\cite{Marczenko:2025lwz} and applied phenomenologically to model neutron-star EOSs and constrain the sound speed in dense QCD matter (see, e.g.~\cite{Moustakidis:2016sab, Margaritis:2019hfq, Laskos-Patkos:2024otk}).

\begin{figure}[t!]
\centering
\includegraphics[width=\columnwidth]{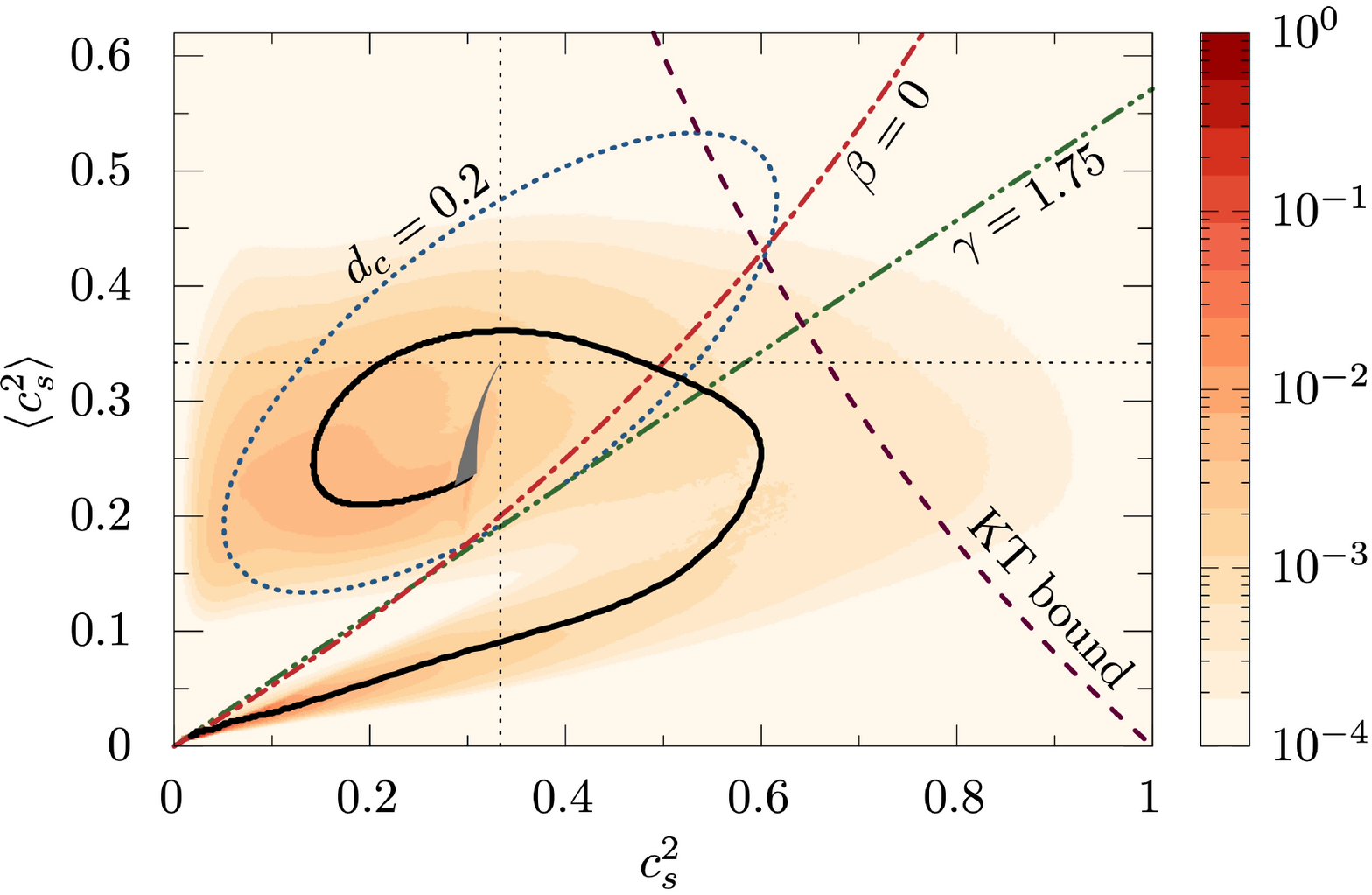}
\caption{Probability density function (PDF) of the EOS in the $\cs-\csavg$ space. The black, solid line shows the averaged EOS obtained by averaging the pressure for given energy density. The arrows on the black line indicate the direction of increasing density. The blue, dotted ellipse shows the $d_c=0.2$ threshold~\citep{Annala:2023cwx}, red, dash-dotted line shows $\beta=0$~\citep{Marczenko:2023txe}, and green, dash-doubly-dotted line marks $\gamma = 1.75$~\citep{Annala:2019puf} threshold. The gray band shows the pQCD constraint. The dashed vertical and horizontal lines mark $\cs = \csavg = 1/3$. The region above the purple, dashed line is not admissible by the kinetic theory (see text for details).}
\label{fig-1}
\end{figure}

The reformulation allows one to represent the KT bound geometrically in the plane spanned by $\cs$ and $\csavg$. This is shown in Fig.~\ref{fig-1}, where the KT bound appears as a curve that excludes a region at large $\cs$ for sufficiently large values of $\csavg$. The bound is inactive near the origin, i.e., low densities, where both $\cs$ and $\csavg$ are small. At low pressures the KT condition smoothly reduces to the causal limit. As the average stiffness increases, the KT bound becomes progressively more restrictive, preventing the coexistence of a large instantaneous sound speed with a large averaged speed of sound. As a result, the constraint disfavors EOSs that attempt to realize extreme stiffening over an extended density interval.

The $\cs-\csavg$ plane can also be used to examine ensembles of EOSs obtained from a statistical ensemble. In Fig.~\ref{fig-1}, we show the probability distribution function (PDF) obtained for a statistical ensemble of EOSs that satisfy the state-of-the-art theoretical and multimessenger constraints. \corr{The ensemble is constructed based on the piecewise-linear speed-of-sound parametrization introduced in~\cite{Annala:2019puf}. We explicitly require that the EOSs are consistent with the chiral effective field theory at low density and perturbative quantum chromodynamics (pQCD) at high density (see~\cite{Marczenko:2022jhl} for details)}. When the KT bound is imposed, a noticeable fraction of this ensemble is excluded, particularly in regions characterized by large $\cs$ combined with moderately large $\csavg$. The KT condition acts as an independent consistency requirement that excludes EOSs otherwise allowed by causality and thermodynamic constraints.

Both the sound speed and its average can be viewed as measures of deviation from conformal behavior, for which $\cs =\csavg=1/3$. Thus, the KT bound constrains how strongly and how persistently the equation of state may depart from conformality. Fig.~\ref{fig-1} indicates commonly used threshold values associated with approximately conformal behavior~\cite{Marczenko:2025hsh, Annala:2023cwx, Marczenko:2023txe}, shown here for comparison with the kinetic-theory bound.

% \section{Conclusion}

To summarize, expressing the kinetic-theory bound in terms of the average sound speed provides a compact way to assess its impact on EOS constructions that are otherwise formulated in terms of integrated stiffness measures. A detailed incorporation of the kinetic-theory constraint into a fully statistical analysis of EOS ensembles is beyond the scope of the present work and will be addressed elsewhere.\\

\section*{Acknowledgment}
This work was partly supported by the program \textit{Excellence Initiative–Research University} of the University of Wroc\l{}aw, funded by the Ministry of Education and Science.


\begin{thebibliography}{}

%\cite{Altiparmak:2022bke}
\bibitem{Altiparmak:2022bke}
S.~Altiparmak, C.~Ecker and L.~Rezzolla,
%``On the Sound Speed in Neutron Stars,''
Astrophys. J. Lett. \textbf{939} (2022) no.2, L34
doi:10.3847/2041-8213/ac9b2a
[arXiv:2203.14974 [astro-ph.HE]].
%220 citations counted in INSPIRE as of 18 Dec 2025
%\cite{Marczenko:2024uit}
\bibitem{Marczenko:2024uit}
M.~Marczenko,
%``Average speed of sound in neutron stars,''
Phys. Rev. C \textbf{110} (2024) no.4, 045811
doi:10.1103/PhysRevC.110.045811
[arXiv:2407.15486 [nucl-th]].
%10 citations counted in INSPIRE as of 18 Dec 2025

%\cite{Marczenko:2025hsh}
\bibitem{Marczenko:2025hsh}
M.~Marczenko,
%``Conformality thresholds in neutron stars,''
J. Subatomic Part. Cosmol. \textbf{3} (2025), 100043
doi:10.1016/j.jspc.2025.100043
[arXiv:2502.10847 [nucl-th]].
%2 citations counted in INSPIRE as of 18 Dec 2025

%\cite{Olson:2000vx}
\bibitem{Olson:2000vx}
T.~S.~Olson,
%``Maximally incompressible neutron star matter,''
Phys. Rev. C \textbf{63} (2000), 015802
doi:10.1103/PhysRevC.63.015802
[arXiv:astro-ph/0011107 [astro-ph]].
%23 citations counted in INSPIRE as of 18 Dec 2025








%\cite{Marczenko:2025lwz}
\bibitem{Marczenko:2025lwz}
M.~Marczenko,
%``Kinetic-Theory Bounds on the Equation of State of Dense QCD Matter,''
[arXiv:2512.17410 [nucl-th]].
%0 citations counted in INSPIRE as of 22 Dec 2025

%\cite{Moustakidis:2016sab}
\bibitem{Moustakidis:2016sab}
C.~C.~Moustakidis, T.~Gaitanos, C.~Margaritis and G.~A.~Lalazissis,
%``Bounds on the speed of sound in dense matter, and neutron star structure,''
Phys. Rev. C \textbf{95} (2017) no.4, 045801
[erratum: Phys. Rev. C \textbf{95} (2017) no.5, 059904]
doi:10.1103/PhysRevC.95.045801
[arXiv:1608.00344 [nucl-th]].
%80 citations counted in INSPIRE as of 20 Dec 2025

%\cite{Margaritis:2019hfq}
\bibitem{Margaritis:2019hfq}
C.~Margaritis, P.~S.~Koliogiannis and C.~C.~Moustakidis,
%``Speed of sound constraints on maximally-rotating neutron stars,''
Phys. Rev. D \textbf{101} (2020) no.4, 043023
doi:10.1103/PhysRevD.101.043023
[arXiv:1910.05767 [nucl-th]].
%26 citations counted in INSPIRE as of 20 Dec 2025

%\cite{Laskos-Patkos:2024otk}
\bibitem{Laskos-Patkos:2024otk}
P.~Laskos-Patkos, G.~A.~Lalazissis, S.~Wang, J.~Meng, P.~Ring and C.~C.~Moustakidis,
%``Speed of sound bounds and first-order phase transitions in compact stars,''
Phys. Rev. C \textbf{111} (2025) no.2, 025801
doi:10.1103/PhysRevC.111.025801
[arXiv:2408.15056 [astro-ph.HE]].
%13 citations counted in INSPIRE as of 20 Dec 2025



%\cite{Marczenko:2022jhl}
\bibitem{Marczenko:2022jhl}
M.~Marczenko, L.~McLerran, K.~Redlich and C.~Sasaki,
%``Reaching percolation and conformal limits in neutron stars,''
Phys. Rev. C \textbf{107} (2023) no.2, 025802
doi:10.1103/PhysRevC.107.025802
[arXiv:2207.13059 [nucl-th]].
%96 citations counted in INSPIRE as of 18 Dec 2025

%\cite{Marczenko:2023txe}
\bibitem{Marczenko:2023txe}
M.~Marczenko, K.~Redlich and C.~Sasaki,
%``Curvature of the energy per particle in neutron stars,''
Phys. Rev. D \textbf{109} (2024) no.4, L041302
doi:10.1103/PhysRevD.109.L041302
[arXiv:2311.13401 [nucl-th]].
%16 citations counted in INSPIRE as of 18 Dec 2025

%\cite{Annala:2023cwx}
\bibitem{Annala:2023cwx}
E.~Annala, T.~Gorda, J.~Hirvonen, O.~Komoltsev, A.~Kurkela, J.~N{\"a}ttil{\"a} and A.~Vuorinen,
%``Strongly interacting matter exhibits deconfined behavior in massive neutron stars,''
Nature Commun. \textbf{14} (2023) no.1, 8451
doi:10.1038/s41467-023-44051-y
[arXiv:2303.11356 [astro-ph.HE]].
%152 citations counted in INSPIRE as of 18 Dec 2025

%\cite{Annala:2019puf}
\bibitem{Annala:2019puf}
E.~Annala, T.~Gorda, A.~Kurkela, J.~N{\"a}ttil{\"a} and A.~Vuorinen,
%``Evidence for quark-matter cores in massive neutron stars,''
Nature Phys. \textbf{16} (2020) no.9, 907-910
doi:10.1038/s41567-020-0914-9
[arXiv:1903.09121 [astro-ph.HE]].
%711 citations counted in INSPIRE as of 18 Dec 2025

\end{thebibliography}
\end{document}